\documentclass[10pt]{article}
\usepackage{sao2}
\usepackage{psfig,epsf}

\issue{2009, 64, 72--105}

\begin{document}
\markboth{Verkhodanov, Verkhodanova, Andernach}
    {Radio identification of decameter-wave sources.
     II:30$\degr$$<\delta<$40$\degr$}
\title{Radio identification of decameter-wave sources. II:
The 30$\degr$$<\delta<$40$\degr$ declination interval}
\author{ O.V. Verkhodanov\inst{a}
    \and N.V. Verkhodanova\inst{a}
    \and H. Andernach\inst{b,c}
}
\institute{
\saoname
\and Argelander-Institut f\"ur Astronomie, Universit\"at Bonn, D-53121, Germany
\and on leave of absence from Departamento de Astronom\'{i}a, Universidad
de Guanajuato, AP 144, Guanajuato CP 36000, Mexico}

\date{July 23, 2008}{August 25, 2008}
\maketitle

\begin{abstract}
This paper is dedicated to the identification of decameter-wave sources 
of the UTR catalog within declination
interval 30$\degr$$<\delta<$40$\degr$. UTR sources are cross-identified 
with CATS database
catalogs within  40'$\times$40' error boxes. The sources are deblended 
using the data on the coordinates
of the objects and the behavior of their continuum radio spectra. The 
spectra of 876 sources are derived
and fitted by standard analytical functions. Of these sources, 221 
objects have straight-line spectra
with spectral indices $\alpha<-1.0$. All objects are catalogued and 
stored in the CATS database.
\\
\mbox{\hspace{1cm}}
\end{abstract}

\section{INTRODUCTION}

The catalog composed by the team of \mbox{Braude et al.} using the 
Ukranian T-shaped Radio Telescope (UTR) in
Khar'kiv and initially published in five
papers~\mbox{\cite{braude1:Verkhodanov_n,braude2:Verkhodanov_n,%
braude3:Verkhodanov_n,braude4:Verkhodanov_n,braude5:Verkhodanov_n},}
contains a total of 1822 radio sources with flux measurements at 10, 
12.6, 14.7, 16.7, 20, and 25 MHz. This
catalog covers 30\% of the sky and is so far the lowest-frequency 
catalog available. Its data can
therefore be used to identify objects in the low-frequency part of the 
spectrum and find their
spectra or estimate the upper decameter-wave flux limits for 
Northern-Sky sources. The original
papers~\mbox{\cite{braude1:Verkhodanov_n,braude2:Verkhodanov_n,%
braude3:Verkhodanov_n,braude4:Verkhodanov_n,braude5:Verkhodanov_n}}
provide no radio identifications for  121 (7\%) sources, whereas most 
of the sources (81\%) lacked
optical identifications.

\begin{figure*}[tpb]
\centerline{
\psfig{figure=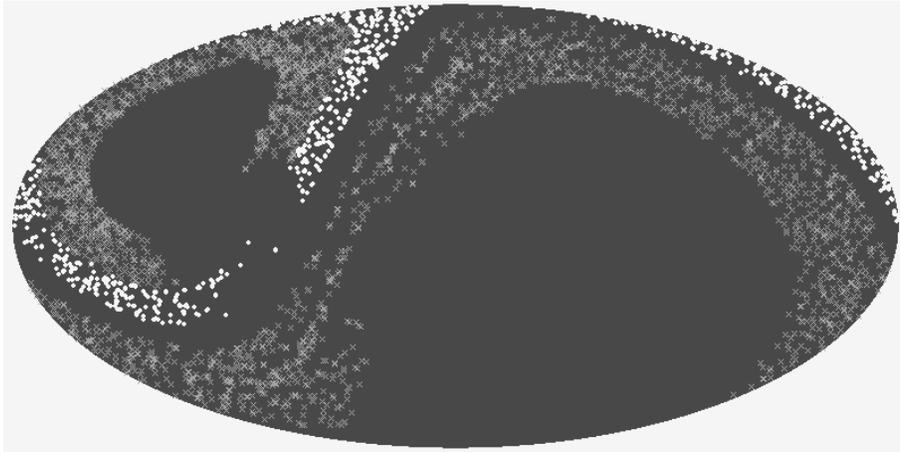,width=12cm,angle=-90}
}
\caption{Sky positions of decameter-wave radio sources  in Galactic 
coordinates. The gray crosses
and white circles indicate the sources from Paper I and those 
identified in this paper, respectively.
}
\end{figure*}

Of special interest are sources with ultrasteep spectra, for which 
low-frequency data points on the
continuum spectra are needed. This subsample contains quite a sizeable 
fraction of distant radio
galaxies~\mbox{\cite{dagkes:Verkhodanov_n,blum_miley:Verkhodanov_n,%
par_big3a:Verkhodanov_n,par_big3b:Verkhodanov_n,uss_list:Verkhodanov_n}.}
The most distant radio galaxies with redshifts $z>4.5$ have been 
discovered among these objects:
$z=5.199$
\cite{debreuk_z:Verkhodanov_n}
and \mbox{$z=4.514$~\cite{kopylov_z:Verkhodanov_n}.}
A~distinctive feature of radio galaxies is that  their parent galaxies
have passively evolving stellar populations even at large redshifts  
($z\sim4$)
\cite{willott_k_z:Verkhodanov_n,verkh_phot_z:Verkhodanov_n},
and this fact can be used to estimate the ages of these stellar 
systems~\mbox{\cite{rg_age:Verkhodanov_n,rg_rc_age:Verkhodanov_n,%
age_sys:Verkhodanov_n}.}
Distant radio galaxies are identified with giant elliptical galaxies 
with absolute magnitudes
$M\sim -26$ whose "central engines" are black holes with masses about 
$10^9M_{\sun}$, which allows them
to be used in cosmological tests in a number of tasks of observational 
radio
cosmology, e.g.,  studies of clustering
\cite{blake_wall:Verkhodanov_n},
``K--$z$'' Hubble diagrams
\cite{willott_k_z:Verkhodanov_n},
\mbox{``size--$z$''
\cite{guerra:Verkhodanov_n,jack_jan:Verkhodanov_n}} and,
``age--$z$'' diagrams
\cite{starob:Verkhodanov_n},
the properties of the radio haloes of clusters of 
galaxies~\cite{colafrancesko:Verkhodanov_n},
etc. (see, e.g., reviews 
\cite{vo_par1:Verkhodanov_n,vo_par2:Verkhodanov_n,%
miley_debreuck:Verkhodanov_n}).

Our decision to perform radio identification of decameter-wave sources
and address the associated deblending problems (i.e., the problems of 
separating the contribution or radio
sources to the total flux measured by a wide-beam radio 
telescope~\cite{deca_nvss:Verkhodanov_n,deca_bul:Verkhodanov_n,deca_aat1
:Verkhodanov_n,deca_aat2:Verkhodanov_n,
deca_id_iau:Verkhodanov_n,deca_id_uss:Verkhodanov_n,%
deca_nat:Verkhodanov_n,deca_azh:Verkhodanov_n}) was inspired, among 
other reasons, by our plans
to compose lists of objects with ultrasteep radio spectra.
A detailed description of the deblending procedure can be found in
\cite{deca_bul:Verkhodanov_n,deca_id_iau:Verkhodanov_n,%
deca_azh:Verkhodanov_n}.
We applied this approach to the first lists of radio sources to obtain 
a catalog
of  2316 objects \cite{deca_bul:Verkhodanov_n} (hereafter referred to 
as Paper~I), which
contribute to the decameter-wave flux, and which could be identified 
with sources at high radio frequencies and in
other wavelength domains, and to find models fits to their radio 
spectra.
We include the identifications from the present paper into the CATS 
database of radio astronomical
catalogs ({\tt http://cats.sao.ru})
\cite{cats:Verkhodanov_n,cats2:Verkhodanov_n}, thereby making them 
publicly available.

Furthermore, supplementing CATS catalogs with decameter-wave sources 
increases the efficiency of: (1)
the analyses of identification lists obtained at different wavelengths 
\mbox{
\cite{irtex1:Verkhodanov_n,irtex2:Verkhodanov_n,irtex3:Verkhodanov_n,%
irtex4:Verkhodanov_n,irtex5:Verkhodanov_n,balayan:Verkhodanov_n,%
trushkin:Verkhodanov_n},}
(2) the search and study of the properties of radio-galaxy
subsamples~\mbox{\cite{rg_gen:Verkhodanov_n,rg_zen0:Verkhodanov_n,%
rg_zen1:Verkhodanov_n,rg_zen2:Verkhodanov_n},}
and
(3) modeling of radio-astronomical surveys performed with 
\mbox{RATAN--600
\cite{compl:Verkhodanov_n,rzf_cmb:Verkhodanov_n,rzf_majorova:Verkhodanov_n}.}
Organization of databases and associating the available heterogeneous 
measurements with
particular objects are tasks of special interest and, as practice 
shows, such compiled catalogs
are popular among the astronomical community
\cite{cats2:Verkhodanov_n,andernach:Verkhodanov_n,vollmer:Verkhodanov_n}.

The publication of a new part of the UTR catalog containing 493 radio 
sources in the declination band \linebreak 
\mbox{30$\degr$$<\delta<$40$\degr$~\cite{braude6:Verkhodanov_n,%
UTR13b:Verkhodanov_n}}
made it possible to expand the list of identifications of such objects
\cite{deca_piter08:Verkhodanov_n}.
The minimum flux of objects in the new UTR catalog is equal to 10\,Jy 
and 20\,Jy at the highest (25\,MHz) and
lowest (10\,MHz) frequencies, respectively. Figure~1 shows sky the 
positions of the new and old objects.

Braude et al.~\cite{braude6:Verkhodanov_n} partially identified their 
objects with low-frequency catalogs and lists
of active galactic nuclei using mostly the 4C 
catalog~\cite{c4:Verkhodanov_n}. We decided to perform a more
detailed identification based on the  CATS database by applying the 
approach described by
Verkhodanov et al.~\cite{deca_bul:Verkhodanov_n}, and to find the 
complete characteristics
of the spectra of deblended decameter-wave objects in the new 
declination interval.

\section{RADIO IDENTIFICATION}

We cross-identified the new list of sources with objects of radio 
catalogs using the same procedure
scheme as we employed in our first paper \cite{deca_bul:Verkhodanov_n}.
Although \mbox{Braude et al.~\cite{braude6:Verkhodanov_n}} report more 
accurate coordinates, we use the
old error boxes in order not to overlook sources within the maximum 
beam size corresponding to the
10-MHz beam. The primary problem
of constructing the spectra of the UTR catalog radio sources found at 
decameter-wave frequencies
(10, 12.6, 14.7, 16.7, 20, and 25\,MHz) was due to the difficulties of 
identifying sources within
large error boxes (within $40'\times40'$ in the case considered) 
originating often from more than one cross-identification
in the CATS database~\cite{cats2:Verkhodanov_n}. To address this 
problem, we performed an interactive reduction of the radio
spectra~\cite{deca_id_iau:Verkhodanov_n} obtained by cross-identifying 
objects of the UTR catalog
with the sources of the CATS database within these $40'$-side boxes. 
Tables~\,1 and \,2 list
the parameters of the main catalogs used at the initial and final 
stages of identification,
respectively.
\begin{table*}
\begin{center}
\caption{
Parameters of the main catalogs used for the identification of 
decameter-wave objects. The MSL (MASTER SOURCE LIST)
contains data compiled from the literature. Data in question were
obtained with radio telescopes operating in
widely different frequency bands and with different angular 
resolutions.}
\begin{tabular}{l|r|c|r|c}
\hline
Name   & Frequency, &  HPBW,      & S$_{lim}$, &  References      \\
      &  MHz    &  arcmin &  Jy      &              \\
\hline
6C     &    151 &  4.2       & $\sim$200 & 
\cite{c6a:Verkhodanov_n,c6b:Verkhodanov_n}   \\
7C     &    151 &  1.2       &     80    & \cite{c7:Verkhodanov_n}      
  \\
MIYUN  &    232 &  3.8       &$\sim$100  & \cite{miyun:Verkhodanov_n}   
  \\
WENSS  &    325 &  0.9       &$\sim$18   & \cite{wenss:Verkhodanov_n}   
  \\
TXS    &    365 &$\sim$0.1   &$\sim$200  & \cite{texas:Verkhodanov_n}   
  \\
B3     &    408 &3$\times$5  &  100      & \cite{b3:Verkhodanov_n}      
  \\
WB92   &   1400 &10$\times$11&  150      & \cite{wb92:Verkhodanov_n}    
  \\
87GB   &   4850 &  3.7       &   25      & \cite{gb87:Verkhodanov_n}    
  \\
GB6    &   4850 &  3.7       &   15      & \cite{gb6:Verkhodanov_n}     
  \\
MSL    &   diff.& diff.      & diff.     & 
\cite{msl1:Verkhodanov_n,msl2:Verkhodanov_n} \\
\hline
\end{tabular} 
\end{center}
\end{table*}

\begin{table*}
\begin{center}
\caption{
Parameters of the catalogs used to refine the coordinates}
\begin{tabular}{l|r|c|r|c}
\hline
Name   & Frequency, &  HPBW,      & S$_{lim}$, &  References      \\
      &  MHz    &  arcmin &  mJy      &              \\
\hline
NVSS   &   1400 &  0.75      &  2.5   & \cite{nvss:Verkhodanov_n}  \\
FIRST  &   1400 &  0.08      &    1   & \cite{first:Verkhodanov_n} \\
VLSS   &     74 &  1.33      &  400   & \cite{vlss:Verkhodanov_n}  \\
\hline
\end{tabular} 
\end{center}
\end{table*}

We used our own software package {\it spg} 
\cite{fadps:Verkhodanov_n,fadps2:Verkhodanov_n} to clean the spectra  in
accordance with the well-developed technique 
\cite{deca_bul:Verkhodanov_n}.
During the cleaning process we removed the sources with spectra that 
did not reach the data points of
the UTR catalog when fitted with standard curves. The search for 
prospective candidates for identification
consists of several steps:
\begin{enumerate}
\item[1)]
    objects of the UTR catalog are cross-identified with the basic 
radio catalogs of the CATS database
with the exception of the highly sensitive (down to 2.5 mJy)  NVSS 
catalog~\cite{nvss:Verkhodanov_n};
\item[2)]
    all objects from the resulting cross-identification list located 
within the  ($40'\times40'$) search
box and having several data points at different frequencies are 
selected;
\item[3)]
    the spectrum of each object is fitted by an appropriate curve and 
extrapolated to the
UTR frequencies;
\item[4)]
    among the objects found inside the search box considered the  radio 
sources are selected
that meet the following conditions:
    \begin{enumerate}
    \item [(a)]
    the flux estimates in the error boxes at the observed UTR 
frequencies must be as close
as possible to the actually observed fluxes; the estimate is obtained
from the value of the fitted spectrum at 16.5 MHz;
    \item [(b)]
    the coordinates of the radio source should be located as close as 
possible to the center of mass
of the coordinates of the UTR sources.
    \end{enumerate}
    There may be as many as five  candidate objects for identification  
(and even in one of the cases)
    (see also Fig.\,2).

\begin{figure*}[tpb]
\centerline{
\vbox{
\hbox{
\psfig{figure=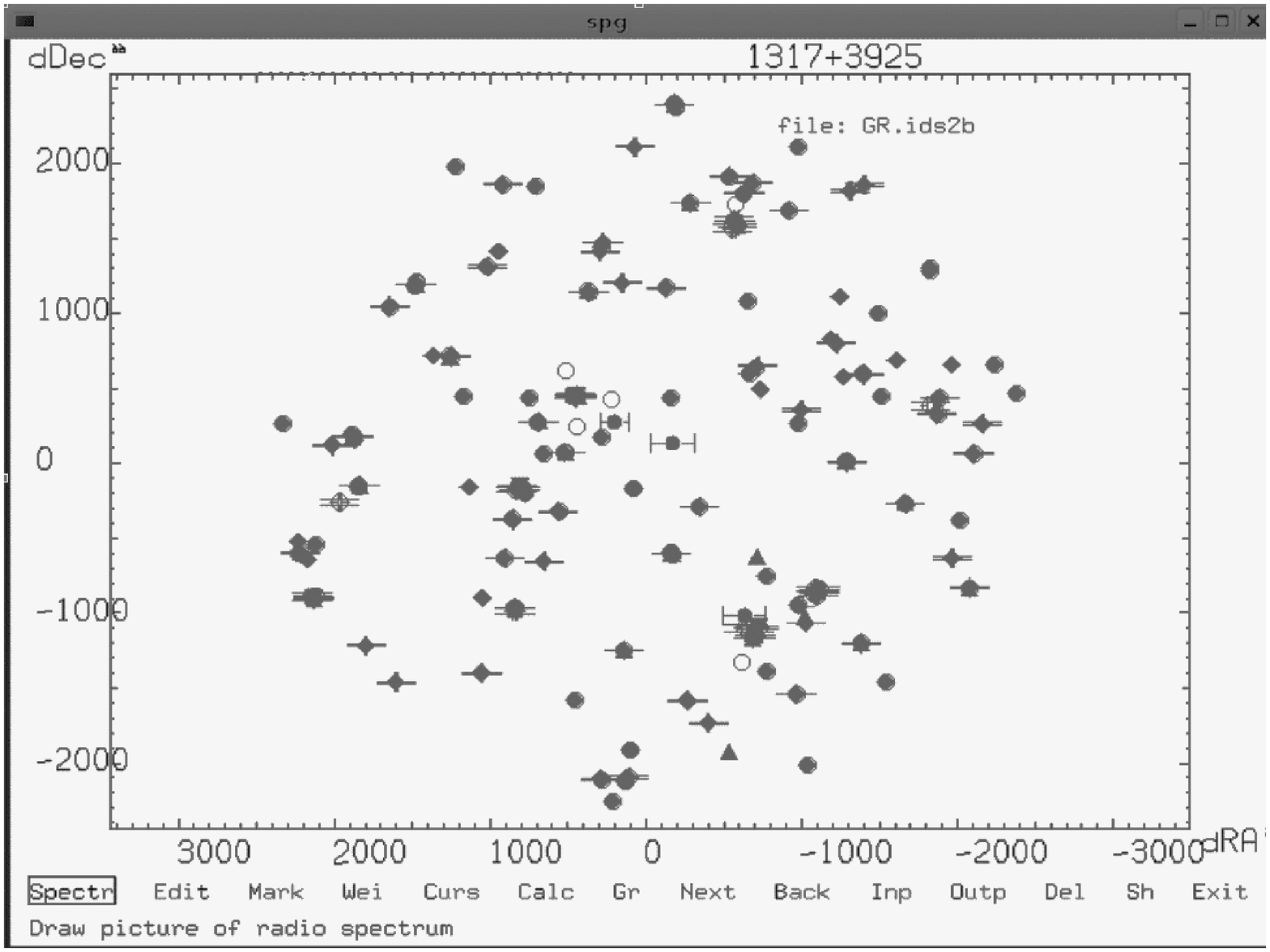,width=8cm,angle=-90}
\psfig{figure=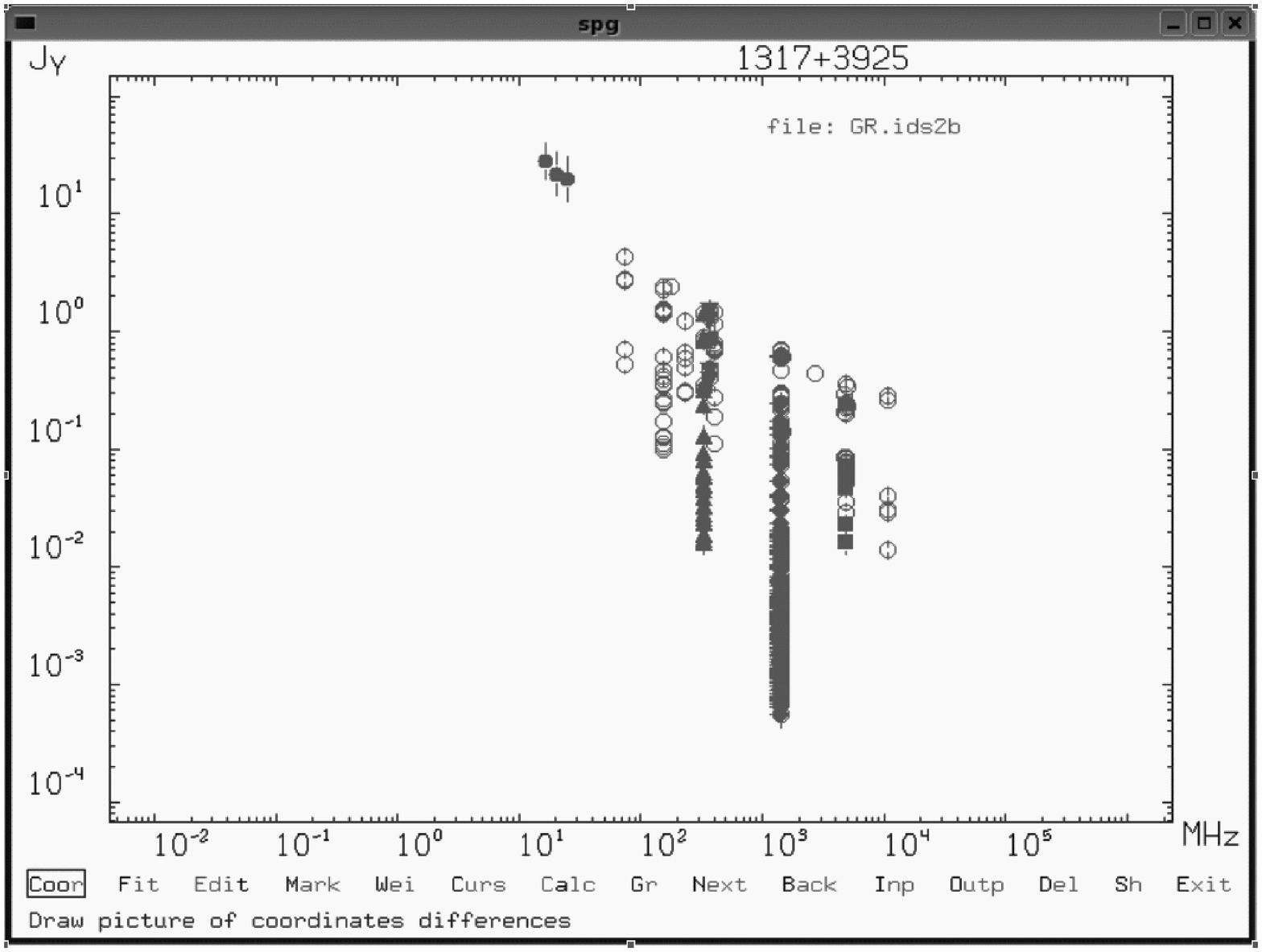,width=8cm,angle=-90}
}
\hbox{
\psfig{figure=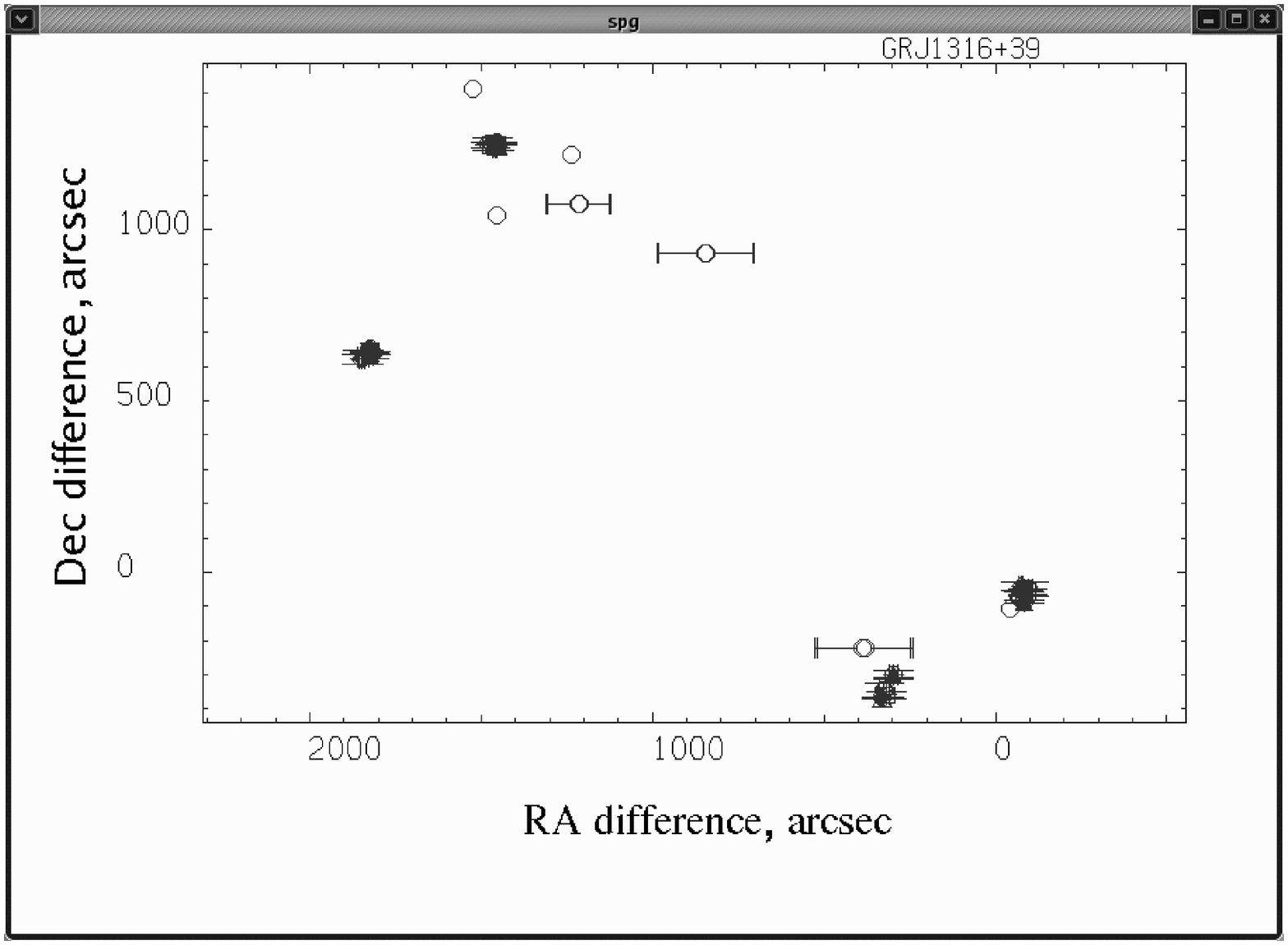,width=8cm,angle=-90}
\psfig{figure=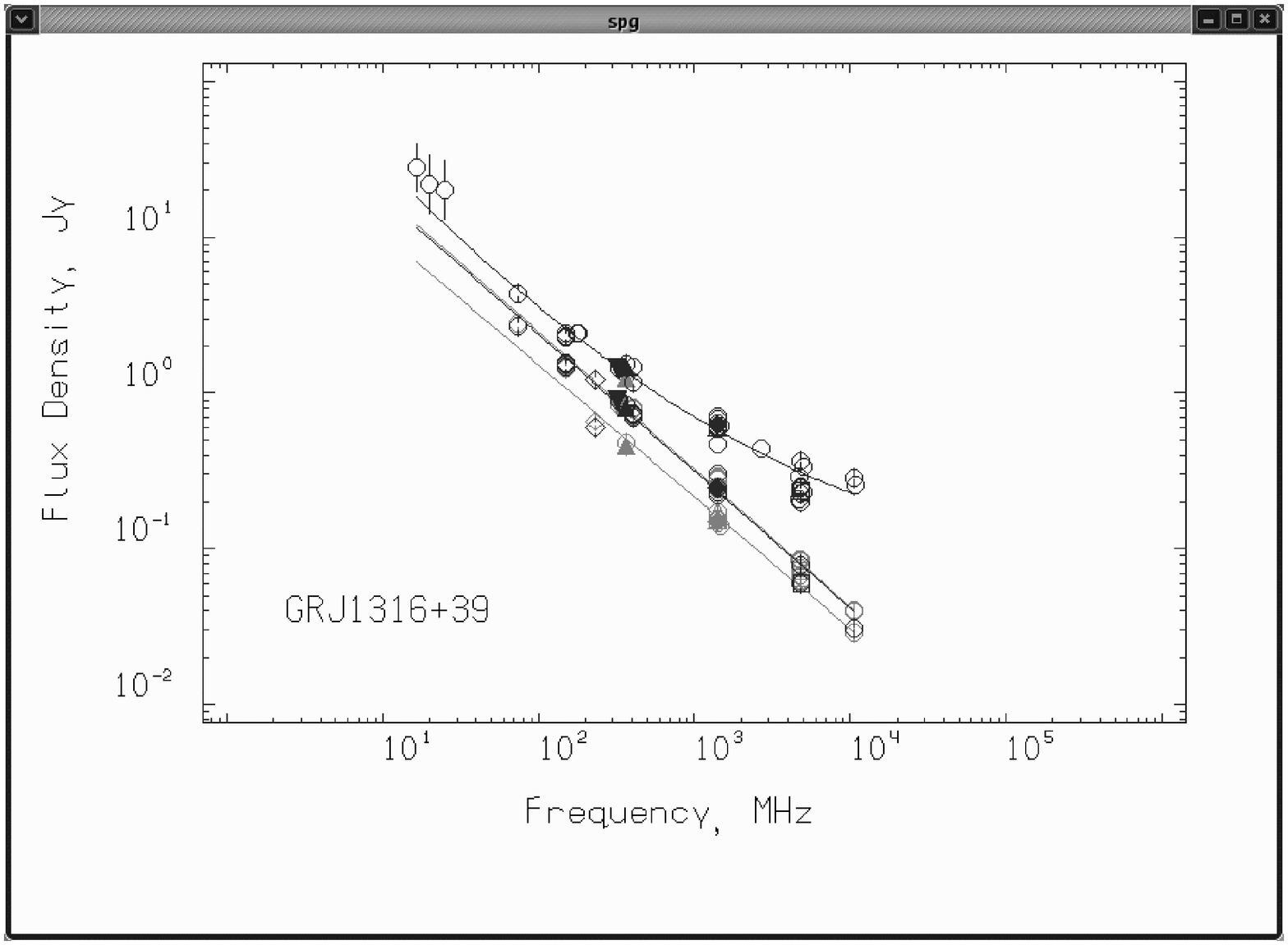,width=8cm,angle=-90}
}}}
\caption{The {\it spg} program screenshot windows showing the initial 
map of radio sources
inside the  $40'\times40'$ search box (top left) and their combined
radio spectrum (top right), as well as the sources in the map, resolved 
into separate objects (bottom left) and
the spectra, resolved in a similar way, fitted with analytical
functions (bottom right). To demonstrate the capabilities of
the {\it spg} program, its working menu is shown in two modes.
}
\end{figure*}

    We assume that all objects meeting the criterion above contribute 
to the UTR source, i.e.,
a blending takes place;
\item[5)]
    we adopt the coordinates of these candidate identifications from 
the following catalogs (in the order of
decreasing priority):
  Texas (365\,MHz)
  \cite{texas:Verkhodanov_n},
  GB6 (4850\,MHz)
  \cite{gb6:Verkhodanov_n}.
  Usually, at least one of these catalogs contains the data on
objects falling within the box;
\item[6)]
  we use the resulting coordinates for cross-identification with NVSS objects
  \cite{nvss:Verkhodanov_n},
  and then use the NVSS data (fluxes) to refine the spectra;
\item[7)]
catalog, if one of its objects
we consider the coordinates listed in the NVSS catalog to be precise, 
if one of its objects
could be cross-identified with the source, otherwise we adopt the 
coordinates obtained as described in
item (5);
\item[8)]
    we use the coordinates obtained to perform cross-identifications 
with non-radio catalogs,
and then use these coordinates for optical identification and study of 
the objects of the UTR catalog.
\end{enumerate}

\noindent

To test the correctness of the spectra obtained, we use the following 
low-frequency catalogs:
6C  (151~MHz)
\cite{c6a:Verkhodanov_n,c6b:Verkhodanov_n},
3C, \mbox{4C (178~MHz)
\cite{c3a:Verkhodanov_n,c3b:Verkhodanov_n,c4:Verkhodanov_n}},
as well as
CL \mbox{(26~MHz)
\cite{cl:Verkhodanov_n},}
\mbox{WKB (38~MHz)
\cite{wkb:Verkhodanov_n},}
and MSH (85~MHz)
\cite{msh1:Verkhodanov_n,msh2:Verkhodanov_n,msh3:Verkhodanov_n}
from the MASTER SOURCE LIST
\cite{msl1:Verkhodanov_n,msl2:Verkhodanov_n}.
Although they did not cover all the previous UTR bands, still they 
demonstrated the
high efficiency of the proposed technique in the regions where 
observational areas
overlapped. The publication
(and hence its  public availability in the CATS database) of the VLSS 
catalog
\cite{vlss:Verkhodanov_n} with declinations $\delta>30\degr$ and with 
measurements at the frequency of 74~MHz was of
great importance for this work; we used it to refine both the radio 
spectra of the sources and their coordinates.

After this identification---and hence refinement of the 
coordinates---we performed a
cross-correlation with the objects of the  NVSS
\cite{nvss:Verkhodanov_n}
and FIRST
\cite{first:Verkhodanov_n},
catalogs whose basic parameters are summarized in Table\,2.

The results of cross-identifications with the data from these catalogs 
allowed us not only to
refine the coordinates, but also to identify multicomponent objects.

\section{THE CATALOG}
\begin{table*}[!tbp]
\begin{center}
\caption{
Statistics of the spectra of radio-identified decameter-wave sources of 
the list studied
(no fits were performed for the spectra of three objects)
}
\begin{tabular}{l|l|r|r}
\hline
 Spectral type & Form of the curve & Number  & \mbox{\hspace*{5mm}}\% \\
\hline
 Straight-line spectrum       & $+A+B*X                      $ &  507  
& 58 \\
 Convex ($C^{+}$)  & $+A {\pm} B*X-C*X^2          $ &   81  &  9 \\
 Concave ($C^{-}$)  & $+A-B*X+C*X^2                $ &  254  & 29 \\
             & ${\pm} A {\pm} B*X+C*EXP(-X) $ &   33  &  4 \\
\hline
\end{tabular}
\end{center}
\end{table*}

We used our identifications of UTR sources to compose a catalog 
containing a total
of 878 objects including all the suspected blends, which is available at
{\tt http://cats.sao.ru/doc/UTR\_ID.html}. The catalog contains for 
each object
its equatorial and Galactic coordinates, the spectral indices at  365, 
1400, and 4850\,MHz, the parameters
of the fitting curve,  the flag indicating whether the object was 
identified with optical,
infrared (IR), or X-ray sources in any of the CATS catalogs. We 
describe the spectra by parametrizing them by the formula
\mbox{$ \lg\,S(\nu) = A + Bx + C f(x)$,} where $S$ is the flux in Jy; 
$x$, the logarithm of frequency  $\nu$
in MHz, and $f(x)$ is one the functions $\exp(-x)$, $\exp(x)$, or $x^2$.
We give the results of identification and the formula used to fit the 
spectra in the
Appendix. The columns give (in this order) the name of the object with 
the blending components
(if any) indicated as $C_i$, the numbers $C_i$ were assigned by source 
power; the equatorial coordinates for the epoch of  J2000.0; the 
Galactic
coordinates; the flag indicating whether the object was identified with 
infrared, optical, and X-ray
sources in any of the CATS catalogs (the symbols used are ``i'', ``o'', 
and ``x'', respectively); the spectral indices at
4850, 1400, and 365~MHz, and the form of the spectrum as defined by the 
above parametrization.

We identified our objects with IR, optical, and X-ray catalogs within 
the boxes of full width with $30''$, $10''$, and $30''$ size,
respectively, centered on the coordinates of the radio
sources. For IR identifications we use the data from the catalogs 2MASS 
and 2MASX
\cite{mass2a:Verkhodanov_n, mass2b:Verkhodanov_n},
\mbox{IFSC
\cite{ifsc:Verkhodanov_n},} and
IPSC
\cite{ipsc:Verkhodanov_n}.
We adopted the optical data from the lists of optical identifications 
of the \mbox{3CR
\cite{c3opt:Verkhodanov_n}}
and Texas
\cite{txso:Verkhodanov_n}
catalogs;
\mbox{SDSS
\cite{sdss:Verkhodanov_n};}
\mbox{VV11
\cite{vv11:Verkhodanov_n};}
APM
\cite{FaAPM:Verkhodanov_n};
UGC
\cite{ugc:Verkhodanov_n};
\mbox{PGC
\cite{pgc:Verkhodanov_n};}
\mbox{USNO
\cite{usno:Verkhodanov_n,usno1:Verkhodanov_n}
etc.}
We searched the X-ray counterparts in the
\mbox{ROSAT
\cite{rosat:Verkhodanov_n},}
1WGA
\cite{wga:Verkhodanov_n},
Einstein
\cite{einst:Verkhodanov_n} and
\mbox{Chandra
\cite{chandra:Verkhodanov_n}} catalogs.
For our decameter-wave sources we found a total of  551 IR, 464 
optical, and 51 X-ray
identifications mostly among active galactic nuclei.

\subsection{Statistics of the Spectra}

The Table\,3 lists the statistics of the spectra of sources fitted by the
standard set of curves
described above. For three objects no fits were performed of the 
spectra due to their complex form.

It is evident from the Table\, 3 that a rather large fraction of 
objects (33\%) exhibits concave
spectra, which may be indicative of the complex structure of the source 
where different
components (the nucleus, hot spots, jets, and the halo) contribute to 
the radio spectrum.

Among the objects with straight-line spectra we found a total of 221 
sources with spectral indices
$\alpha < -1.0$ ($S\sim\nu^\alpha$). This subsample also includes 98 
objects with ultrasteep
spectra ($\alpha < -1.1$).

\begin{figure}[tpb]
\centerline{
\psfig{figure=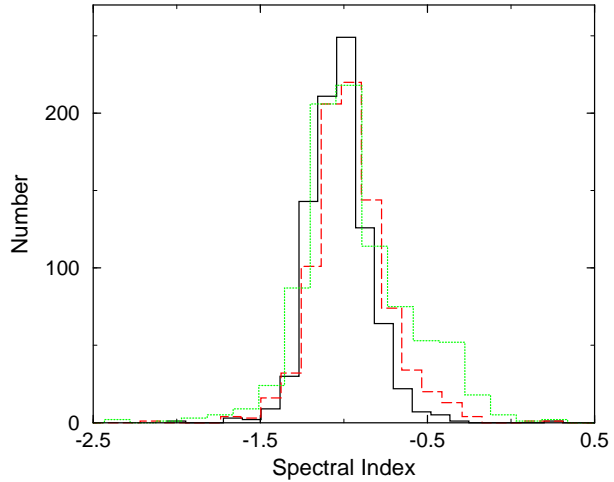,width=8cm,angle=-90}
}
\caption{
Distribution of the spectral indices of decameter-wave sources at 365 
(the solid line), 1400
(the dashed line), and 4850~MHz (the dotted line).
}
\end{figure}

\begin{figure}[tpb]
\centerline{
\psfig{figure=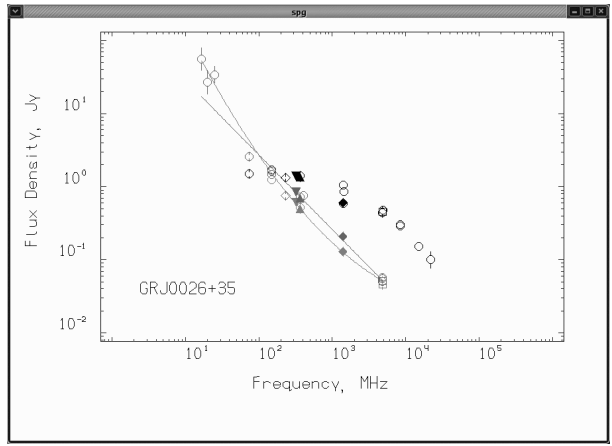,width=8cm,angle=-90}
}
\caption{
The spectra of the blending components of the GRJ0026+35 radio source. 
The spectrum of the central
object (markers aligned along the convex spectrum) is not fitted and 
shows flat dependence  $\log S(\log \nu)$
at low frequencies. Different symbols correspond to different catalogs.
}
\end{figure}

\begin{figure}[tpb]
\centerline{
\psfig{figure=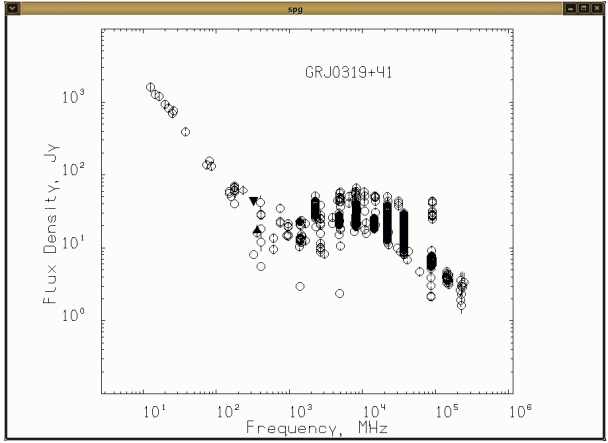,width=8cm,angle=-90}
}
\caption{
Spectra of the GRJ0319+41 radio galaxy (Perseus~A). The fluxes measured
at different epochs show substantial scatter at high frequencies; this 
is due to its known variability.
}
\end{figure}
We used the models of the spectra to build the histograms of spectral 
indices (Fig.\,3). We compute
the spectral index as the slope of the spectrum (in double-logarithmic 
scale) at the given frequency.
The distribution of the 4850-MHz indices in Fig.\,3 shows an excess of 
objects in the
right-hand side of the diagram, which is due to the excess of sources 
with concave spectra,
which, in turn, must be due to the contribution of the flux from the 
nuclear component
of the object.

\subsection{Some Remarkable Sources}

\begin{figure}[!t]
\centerline{
\psfig{figure=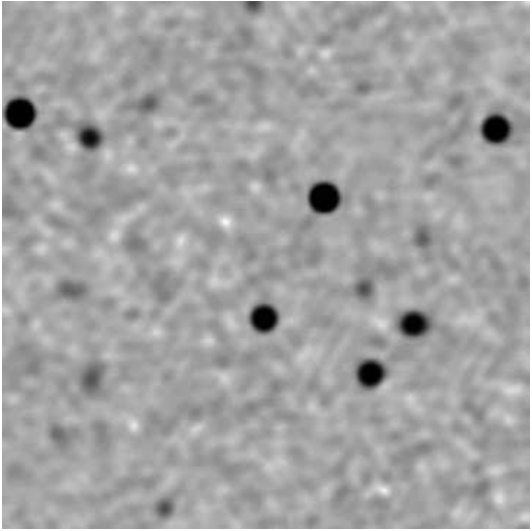,width=7cm}
}
\caption{
The NVSS map of the $30'\times 30'$ region centered on GRJ1646+40. The 
dark spots are
radio sources.
}
\end{figure}
We now point out several interesting objects included in our list.

We did not perform fitting procedures for the three radio sources 
GRJ0026+35, GRJ0319+41,
and GRJ0555+40($C_4$) because of the evidently multicomponent nature of 
their spectra.

GRJ0026+35 is blended with at least two other sources, and the object 
that is the closest to the
coordinates mass center of the decameter-wave measurements shows a 
low-frequency
flattening in its spectrum (Fig.\,4). The source is rather powerful at 
high frequencies and may
have a halo emitting at low frequencies, which prevents the cutoff in 
the spectrum.

GRJ0319+41 is the well-studied source \linebreak Perseus A (3C~84, 
NGC~1275). Its high-frequency radio spectrum
shows very large scatter due to variability (Fig.\,5).

GRJ0555+40($C_4$) is a powerful object at low frequencies. It is 
located in the field of the
decameter-wave source. GRJ0555+40 shows a cutoff at low frequencies and 
it most likely does not
contribute to the decameter-wave flux.

The UTR radio source GRJ1646+40 has the maximum number of blending 
components among the objects
of our list (Fig.\,6). Despite the apparently high density of bright 
radio objects, the effect is
most likely due to chance alignment.

We also point out the radio sources with spectral indices smaller than 
$-1.3$:
\begin{itemize}

\item   GRJ0154+40($C_2$)  with $\alpha=-1.44$;
\item   GRJ0555+40($C_2$)  with $\alpha=-1.57$;
\item   GRJ1055+37         with $\alpha=-1.66$;
\item   GRJ1153+37         with $\alpha=-1.66$; the source coincides with
the galaxy cluster MaxBCG \linebreak J178.15191+37.54548 in NED;
\item   GRJ1240+37($C_4$)  with $\alpha=-1.42$; the source coincides with
the galaxy cluster MaxBCG \linebreak J189.97692+37.73205 in NED;
\item   GRJ2313+38         with $\alpha=-1.48$; the source is probably 
a cluster dominated by the  2MASX galaxy J23134832+3842315.
\end{itemize}

\section{CONCLUSIONS}

We performed cross-identification of decameter-wave radio sources of 
the UTR catalog
in the declination interval 30$\degr$$<\delta<$40$\degr$ by searching 
for objects from
the catalogs of the CATS database within the  $40'\times40'$ error box. 
We built
the spectra of 876 sources and fitted them by standard analytical
functions.
These sources include 221 objects with straight-line spectra and 
spectral indices $\alpha<-1.0$.
We provide access to all our catalogued objects from within the CATS 
database.

\begin{acknowledgements}
This work makes use if the CATS database of radio-astronomical 
catalogs\footnote{\tt http://cats.sao.ru}
\cite{cats:Verkhodanov_n,cats2:Verkhodanov_n}
and FADPS system of reduction of radio astronomical data
\footnote{\tt http://sed.sao.ru/$\sim$vo/fadps\_e.html}
\mbox{\cite{fadps:Verkhodanov_n,fadps2:Verkhodanov_n}.}
Authors are thankful to S.A.~Trushkin for the compilation of the 
initial UTR~XIII catalog within the
CATS database. H.A. acknowledges support from grant 81356 from CONACyT 
of Mexico and the hospitality of
the Emmy-Noether Research Group of T.~Reiprich at AIfA Univ. of Bonn, 
Germany, where partial support from
the Transregional Collaborative Research Center TRR33 ``The Dark 
Universe'' was received.
This work was supported partially by the Russian Foundation for Basic 
Research (grants 07-02-01417, 08-02-00159, and 08-02-00504).
\end{acknowledgements}

\clearpage

\newcommand {\hs}{\hspace*{1.5mm}}   
\newcommand {\hd}{\hspace*{3.0mm}}   
\newcommand {\ha}{\hspace*{0.75mm}}  
\newcommand {\hn}{\hspace*{2.25mm}}  
\newcommand {\ho}{\hspace*{3.75mm}}  
\newcommand {\hp}{\hspace*{5.25mm}}  

\onecolumn
{
\small
\topcaption {APPENDIX. Radio identification and spectra of decametric
sources.}
\tablefirsthead{
  \hline \multicolumn{1}{|c}{Name}                 
    & \multicolumn{1}{|c|}{RA+Dec (J2000.0)}       
    & \multicolumn{1}{|c|}{$l+b$}                  
    & \multicolumn{1}{|c|}{Id}                     
    & \multicolumn{1}{|c|}{$\alpha$}               
    & \multicolumn{1}{|c|}{Spectrum}          \\   
  \hline \multicolumn{1}{|c}{}                     
    & \multicolumn{1}{|c|}{hhmmss.ss+ddmmss.s}     
    & \multicolumn{1}{|c|}{$^\circ~~~~^\circ$}     
    & \multicolumn{1}{|c|}{}                       
    & \multicolumn{1}{|c|}{4850,1400,365}          
    & \multicolumn{1}{|c|}{}                  \\   
  \hline \multicolumn{1}{|c}{1}                    
    & \multicolumn{1}{|c|}{2}                      
    & \multicolumn{1}{|c|}{3}                      
    & \multicolumn{1}{|c|}{4}                      
    & \multicolumn{1}{|c|}{5}                      
    & \multicolumn{1}{|c|}{6}                 \\   
  \hline
}
\tablehead {\hline
  \hline \multicolumn{1}{|c}{1}                    
    & \multicolumn{1}{|c|}{2}                      
    & \multicolumn{1}{|c|}{3}                      
    & \multicolumn{1}{|c|}{4}                      
    & \multicolumn{1}{|c|}{5}                      
    & \multicolumn{1}{|c|}{6}                 \\   
  \hline \multicolumn{1}{|c}{Name}                 
    & \multicolumn{1}{|c|}{hhmmss.ss+ddmmss.s}     
    & \multicolumn{1}{|c|}{$l^\circ~+~b^\circ$}    
    & \multicolumn{1}{|c|}{Id}                     
    & \multicolumn{1}{|c|}{$\alpha$}               
    & \multicolumn{1}{|c|}{Spectrum}          \\   
  \hline
}
\tabletail{\hline}

\begin{center}
\begin{supertabular}{|lccccl|}
\hline
GRJ0013+34  $   $ & 001344.09$+$344140.6 & 114.844$-$27.348 & io  &$-$0.80 $-$0.80 $-$0.80&$+2.648-0.802x              $ \\
GRJ0013+40  $   $ & 001331.09$+$405135.9 & 115.812$-$21.245 & iox &$-$0.28 $-$0.50 $-$0.75&$+3.852-1.818x+0.209x^2     $ \\
GRJ0013+40  $C_2$ & 001252.90$+$403246.6 & 115.634$-$21.538 &     &$-$0.67 $-$0.88 $-$1.11&$+4.147-2.120x+0.197x^2     $ \\
GRJ0013+40  $C_3$ & 001313.35$+$405300.4 & 115.756$-$21.214 & i   &$-$0.28 $-$0.71 $-$1.18&$+5.268-3.240x+0.402x^2     $ \\
GRJ0014+32  $   $ & 001506.23$+$321613.2 & 114.746$-$29.787 & io  &$-$0.53 $-$0.60 $-$0.68&$+2.809-1.035x+0.069x^2     $ \\
GRJ0014+32  $C_2$ & 001613.30$+$323839.3 & 115.078$-$29.452 & io  &$-$0.76 $-$0.88 $-$1.03&$+3.622-1.640x+0.120x^2     $ \\
GRJ0026+35  $   $ & 002641.64$+$350842.5 & 117.873$-$27.233 & io  &$-$0.43 $-$0.48 $-$0.57&$ halo~?                    $ \\
GRJ0026+35  $C_2$ & 002652.01$+$355623.9 & 117.998$-$26.445 & io  &$-$0.58 $-$0.86 $-$1.16&$+4.392-2.502x+0.261x^2     $ \\
\hline
\end{supertabular}
\end{center}
}


\begin{thebibliography}{99}

\bibitem{braude1:Verkhodanov_n}
1.
S.~Ya.~Braude, A.~V.~Megn, S.~L.~Rashkovski, et al.,
     \apss~ {\bf 54}, 37 (1978).
\bibitem{braude2:Verkhodanov_n}
2.
S.~Ya.~Braude, A.~V.~Megn, K.~P.~Sokolov, et al.,
     \apss~ {\bf 64}, 73 (1979).
\bibitem{braude3:Verkhodanov_n}
3.
S.~Ya.~Braude, A.~P.~Miroshnichenko, K.~P.~Sokolov, and N.~K.~Sharykin,
    \apss~ {\bf 74}, 409 (1981).
\bibitem{braude4:Verkhodanov_n}
4.
S.~Ya.~Braude,  N.~K.~Sharykin, K.~P.~Sokolov, and S.~M.~Zakharenko,
     \apss~ {\bf 111}, 1 (1985).
\bibitem{braude5:Verkhodanov_n}
5.
 S.~Ya.~Braude, K.~P.~Sokolov, and S.~M.~Zakharenko, \apss~
     {\bf 213}, 1 (1994).

\bibitem{dagkes:Verkhodanov_n}
6.
  R.~D.~Dagkesamanskii,
    Nature {\bf 226}, 432 (1970).
\bibitem{blum_miley:Verkhodanov_n}
7.
  G.~Blumenthal and G.~Miley, \aaa~ {\bf 80}, 13 (1979).
\bibitem{uss_list:Verkhodanov_n}
8.
  C.~de~Breuck,   W.~van~Breugel,  H.~J.~A.~R\"ottgering, and G.~Miley,
   \aas~ {\bf 143}, 303 (2000).
\bibitem{par_big3a:Verkhodanov_n}
9.
 Yu.~N.~Parijskij,  W.~M.~Goss,  A.~I.~Kopylov, et al.,
    \bsao~ {\bf 40}, 5 (1996).
\bibitem{par_big3b:Verkhodanov_n}
10.
 Yu.~N.~Parijskij, W.~M.~Goss, A.~I.~Kopylov, et al.,
      Astron. Astrophys. Trans. {\bf 18}, 437 (1999).

\bibitem{debreuk_z:Verkhodanov_n}
11.
  De Breuck C., W.~van~Breugel,  S.~A.~Stanford, et al.,
  \aj~  {\bf 123}, 637 (2002).
\bibitem{kopylov_z:Verkhodanov_n}
12.
 A.~I.~Kopylov, W.~M.~Goss, Yu.~N.~Pariiskii, et al.,
     Astronom. Lett. {\bf 32}, 433 (2006), arXiv:0705.2971.
\bibitem{willott_k_z:Verkhodanov_n}
13.
 C.~J.~Willot,  R.~J.~McLure, and  M.~J.~Jarvis,
   \apj~ {\bf 587}, L1  (2003).
\bibitem{verkh_phot_z:Verkhodanov_n}
14.
 O.~V.~Verkhodanov, A.~I.~Kopylov, Yu.~N.~Pariiskii, et al.,
 Astronomy Lett. {\bf 31}, 221 (2005), arXiv:0705.3046.
\bibitem{rg_age:Verkhodanov_n}
15.
O.~V.~Verkhodanov; A.~I.~Kopylov, Yu.~N.~Parijskij, et al.,
   \bsao~ {\bf 48},  41 (1999), astro-ph/9910559.
\bibitem{rg_rc_age:Verkhodanov_n}
16.
O.~V.~Verkhodanov,  Yu.~N.~Parijskij, N.~S.~Soboleva, et al.,
     \bsao~ {\bf 52},  5 (2001), astro-ph/0203522.
\bibitem{age_sys:Verkhodanov_n}
17.
O.~V.~Verkhodanov, A.~I.~Kopylov, O.~P.~Zhelenkova, et al.,
     Atsron. Astrophys. Trans. {\bf 19}, 663 (2000)
     astro-ph/9912359.

\bibitem{blake_wall:Verkhodanov_n}
18.
 C.~Blake and J.~Wall,
    \mnras~ {\bf 337}, 993 (2002).
\bibitem{guerra:Verkhodanov_n}
19.
 E.~J.~Guerra,  R.~A.~Daly, and L.~Wan,
  \apj~ {\bf 544}, 659 (2000).
\bibitem{jack_jan:Verkhodanov_n}
20.
 J.~C.~Jackson and  A.~L.~Jannetta,
  JCAP {\bf 11}, 002 (2006).
\bibitem{starob:Verkhodanov_n}
21.
 O.~V.~Verkhodanov,  Yu.~N.~Parijskij and  A.~A.~Starobinsky,
    \bsao~ {\bf 58}, 5 (2005), arXiv:0705.2776.
\bibitem{colafrancesko:Verkhodanov_n}
22.
S.~Colafrancesco and B.~Mele,
    \apj~ {\bf 562}, 24  (2001).

\bibitem{vo_par1:Verkhodanov_n}
23.
 O.~V.~Verkhodanov and  Yu.~N.~Parijskij,
    \bsao~ {\bf 55}, 66 (2003).
\bibitem{vo_par2:Verkhodanov_n}
24.
  O.~V.~Verkhodanov and  Yu.~N.~Parijskij,
    Radio galaxies and Cosmology,
    (Fiz.Mat.Lit., Moscow 2008) [in Russian] (in press).
\bibitem{miley_debreuck:Verkhodanov_n}
25.
G.~Miley and C.~De~Breuck,
  \aar~ {\bf 15}, 67 (2008).


\bibitem{deca_nvss:Verkhodanov_n}
26.
  O.~Verkhodanov, H.~Andernach, N.~Verkhodanova and N.~Loiseau,
    in {\it  Proceedings of  ``Observational Cosmology with the New 
Radio Surveys''},
     Ed. by M.~Bremer, N.~Jackson and I.~P\'erez-Fournon,
     Kluwer Acad. Publ., ASSL {\bf 226}, 255 (1998), astro-ph/9703142.
\bibitem{deca_bul:Verkhodanov_n}
27.
   O.~Verkhodanov, H.~Andernach and N.~Verkhodanova,
    \bsao~ {\bf 49}, 53 (2000). astro-ph/0008431.
\bibitem{deca_aat1:Verkhodanov_n}
28.
O.~V.~Verkhodanov, H.~Andernach and N.~V.~Verkhodanova,
     Atsron. Astrophys. Trans. {\bf 19}, 543 (2000).
\bibitem{deca_aat2:Verkhodanov_n}
29.
O.~V.~Verkhodanov, H.~Andernach and N.~V.~Verkhodanova,
      Astron. Astrophys. Trans. {\bf 20}, 321 (2001).
\bibitem{deca_id_iau:Verkhodanov_n}
30.
O.~V.~Verkhodanov, H.~Andernach and N.~V.~Verkhodanova,
     in {\it Proceedings of 199 Symp. IAU },
      {\it ``The Universe at Low Radio Frequencies''},
     Ed. by A.~Pramesh Rao, G.~Swarup, and Gopal-Krishna, p.215 (2002),
     astro-ph/9912331.
\bibitem{deca_id_uss:Verkhodanov_n}
31.
H.~Andernach, O.~V.~Verkhodanov and N.~V.~Verkhodanova,
     in {\it Proceedings of 199 Symp. IAU},
    {\it ``The Universe at Low Radio Frequencies''},
     Ed. by A.~Pramesh Rao, G.~Swarup, and Gopal-Krishna, 217 (2002).
\bibitem{deca_nat:Verkhodanov_n}
32.
N.~V.~Verkhodanova, O.~V.~Verkhodanov, and H.~Andernach,
   in {\it Proceedings of  184 Colloq. IAU},
  {\it `` AGN Surveys''},
   Ed. by  R.~F.~Green, E.~Ye.~Khachikian, and D.~B.~Sanders,
   ASP Conf. Ser. {\bf 284}, 306 (2002), astro-ph/0112143.
\bibitem{deca_azh:Verkhodanov_n}
33.
 O.~V.~Verkhodanov,  N.~V.~Verkhodanova, and H.~Andernach,
      Astronomy Reports {\bf 47}, 110 (2003)


\bibitem{cats:Verkhodanov_n}
34.
O.~V.~Verkhodanov, S.~A.~Trushkin, H.~Andernach, and V.~N.~Chernenkov,
      ASP Conf. Ser.  {\bf 125}, 322 (1997).
\bibitem{cats2:Verkhodanov_n}
35.
O.~V.~Verkhodanov, S.~A.~Trushkin, H.~Andernach, and V.~N.~Chernenkov,
    \bsao~ {\bf 58}, 118 (2005), arXiv:0705.2959.

\bibitem{irtex1:Verkhodanov_n}
36.
 S.~A.~Trushkin and O.~V.~Verkhodanov,
   \bsao~ {\bf 39}, 150 (1995).
\bibitem{irtex2:Verkhodanov_n}
37.
 S.~A.~Trushkin and  O.~V.~Verkhodanov,
     Baltic Astronomy  {\bf 6},  345 (1997).
\bibitem{irtex3:Verkhodanov_n}
38.
O.~V.~Verkhodanov and S.~A.~Trushkin,
     \bsao~ {\bf 50}, 115 (2000).
\bibitem{irtex4:Verkhodanov_n}
39.
O.~V.~Verkhodanov, V.~H.~Chavushyan, R.~Mujica, et al.,
     Astron. Reports  {\bf 47}, 119 (2003).
\bibitem{irtex5:Verkhodanov_n}
40.
V.~H.~Chavushyan, O.~V.~Verkhodanov, J.~R.~Valdes, et al.,
     Astrophysics {\bf 48}, 113 (2005).
\bibitem{balayan:Verkhodanov_n}
41.
 S.~K.~Balayan and  O.~V.~Verkhodanov,
Survey.
     Astrophysics {\bf 47}, 505 (2004).
\bibitem{trushkin:Verkhodanov_n}
42.
   S.~A.~Trushkin,
   \bsao~ {\bf 55}, 90 (2003).

\bibitem{rg_gen:Verkhodanov_n}
43.
 O.~V.~Verkhodanov,  A.~I.~Kopylov,  Yu.~N.~Parijskij, et al.,
   \bsao~ {\bf  48}, 41 (1999), astro-ph/9910559.

\bibitem{rg_zen0:Verkhodanov_n}
44.
 O.~V.~Verkhodanov,
     Astronomy Reports {\bf 38}, 307 (1994).

\bibitem{rg_zen1:Verkhodanov_n}
45.
 O.~V.~Verkhodanov and  N.~V.~Verkhodanova,
     Astron. Reports {\bf 43}, 417 (1999).
\bibitem{rg_zen2:Verkhodanov_n}
46.
Yu.~N.~Parijskij, W.~M.~Goss, O.~V.~Verkhodanov, et al.,
    \bsao~ {\bf 48},  5 (1999), astro-ph/9910383.



\bibitem{compl:Verkhodanov_n}
47.
 V.~L.~Gorokhov and  O.~V.~Verkhodanov,
     Astronomy Lett. {\bf 20}, 671 (1994).
\bibitem{rzf_cmb:Verkhodanov_n}
48.
M.~L.~Khabibullina, O.~V.~Verkhodanov, and Yu.~N.~Parijskij,
   \ab~ {\bf 63}, 101 (2008).
\bibitem{rzf_majorova:Verkhodanov_n}
49.
E.~K.~Majorova, 
  \ab~ {\bf 63}, 56 (2008).

\bibitem{andernach:Verkhodanov_n}
50.
H.~Andernach,
   Astrophys. Lett. Commun. {\bf 31},  1 (1995).
\bibitem{vollmer:Verkhodanov_n}
51.
 B.~Vollmer, E.~Davoust, P.~Dubois, et al.,
 \aaa~ {\bf 431}, 1177 (2005).


\bibitem{braude6:Verkhodanov_n}
52.
 S.~Ya.~Braude et al.,
     \apss~ {\bf 280}, 235 (2002).
\bibitem{UTR13b:Verkhodanov_n}
53.
S.~Ya.~Braude et al., Kinematika i Fizika Nebesnykh Tel, {\bf 19}, 291 
(2003).

\bibitem{deca_piter08:Verkhodanov_n}
54.
 O.~V.~Verkhodanov,  N.~V.~Verkhodanova, H.~Andernach,
     in {\it Proceedings of Internat. Conf. ``Problems of Practical 
Cosmology''},
     Ed. by Yu.~Baryshev, Igor~ N.~Taganov, P.~Teerikorpi,
     Russian Geograph. Soc., St.Petersburg, V.{\bf II}, 251 (2008).

\bibitem{c4:Verkhodanov_n}
55.
   J.~D.~H.~Pilkington and P.~F.~Scott, Mem. Roy. Astron. Soc. {\bf 69},
     183 (1965).

\bibitem{c6a:Verkhodanov_n}
56.
  S.~E.~G.~Hales, J.~E.~Baldwin, and P.~J.~Warner, \mnras~ {\bf 234}, 
919 (1988).
\bibitem{c6b:Verkhodanov_n}
57.
  S.~E.~G.~Hales,  C.~R.~Masson,  P.~J.~Warner, and  J.~E.~Baldwin,
     \mnras~ {\bf 246}, 256 (1990).
\bibitem{c7:Verkhodanov_n}
58.
   M.~M.~McGilchrist,  J.~E.~Baldwin,  J.~M.~Riley, et al.,
     \mnras~ {\bf 246}, 110  (1990).
\bibitem{miyun:Verkhodanov_n}
59.
    X.~Zhang,  Y.~Zheng,  H.~Chen, et al., \aas~ {\bf 121}, 59 (1997).
\bibitem{wenss:Verkhodanov_n}
60.
    R.~B.~Rengelink,  Y.~Tang,  A.~G.~de~Bruyn, et al.,
       \aas~ {\bf 124}, 259 (1997).
\bibitem{texas:Verkhodanov_n}
61.
   J.~N.~Douglas, F.~N.~Bash,  F.~A.~Bozyan, et al., 
     \aj~ {\bf 111}, 1945  (1996).
\bibitem{b3:Verkhodanov_n}
62.
    A.~Ficarra,  G.~Grueff, and  G.~Tomassetti, \aas~ {\bf 59}, 255 
(1985).
\bibitem{wb92:Verkhodanov_n}
63.
   R.~L.~White and  R.~H.~Becker, \apjs~  {\bf 79}, 331 (1992).
\bibitem{gb87:Verkhodanov_n}
64.
      P.~C.~Gregory and  J.~J.~Condon, \apjs~  {\bf 75}, 1011 (1991).
\bibitem{gb6:Verkhodanov_n}
65.
   P.~C.~Gregory,  W.~K.~Scott, K.~Douglas, and J.~J.~Condon,
     \apjs~ {\bf 103}, 427 (1996).
\bibitem{msl1:Verkhodanov_n}
66.
    R.~S.~Dixon, \apjs~  {\bf 20}, 1 (1970).
\bibitem{msl2:Verkhodanov_n}
67.
    R.~S.~Dixon, 1981, Master List of Radio Sources, version 43
   (with corrections by H.\,Andernach, 15/Nov/98).

\bibitem{nvss:Verkhodanov_n}
68.
   J.~J.~Condon,  W.~D.~Cotton,  E.~W.~Greisen, et al.,
     \aj {\bf 115}, 1693 (1998).
\bibitem{first:Verkhodanov_n}
69.
  R.~L.~White,  R.~H.~Becker,  D.~J.~Helfand, and  M.~D.~Gregg, \apj~
    {\bf 475}, 479  (1997).
\bibitem{vlss:Verkhodanov_n}
70.
A.~S.~Cohen, W.~M.~Lane,  W.~D.~Cotton, et al.,
   \aj~ {\bf 134}, 1245 (2007).

\bibitem{fadps:Verkhodanov_n}
71.
O.~V.~Verkhodanov,
      ASP Conf. Ser. {\bf 125}, 46 (1997).
\bibitem{fadps2:Verkhodanov_n}
72.
O.~V.~Verkhodanov, B.~L.~Erukhimov, M.~L.~Monosov, et al.,
    \bsao~ {\bf 36}, 132 (1993).

\bibitem{c3a:Verkhodanov_n}
73.
 D.~O.~Edge,  J.~R.~Shakeshaft, W.~B.~McAdam, et al., 
J.E.,Archer S.,
     \mnras~ {\bf 68}, 37 (1959).
\bibitem{c3b:Verkhodanov_n}
74.
   A.~S.~Bennett, Mem. Roy. Astr. Soc. {\bf 68}, 163  (1961).
\bibitem{cl:Verkhodanov_n}
75.
   M.~R.~Viner and W.~C.~Erickson,  \aj~ {\bf 80}, 931 (1975).
\bibitem{wkb:Verkhodanov_n}
76.
   P.~J.~S.~Williams,  S.~Kenderdine, and  J.~E.~Baldwin,
    Mem. R. Astron. Soc. {\bf 70}, 53 (1966).
\bibitem{msh1:Verkhodanov_n}
77.
  B.~Y.~Mills, O.~B.~Slee, and E.~R.~Hill, Australian J. Phys. {\bf11}, 360
	(1958).
\bibitem{msh2:Verkhodanov_n}
78.
  B.~Y.~Mills, O.~B.~Slee, and E.~R.~Hill, Australian J. Phys. {\bf 13}, 676
	 (1960).
\bibitem{msh3:Verkhodanov_n}
79.
  B.~Y.~Mills, O.~B.~Slee, and E.~R.~Hill, Australian J. Phys. {\bf 14}, 497
	 (1961).


\bibitem{mass2a:Verkhodanov_n}
80.
M.~F.~Skrutskie, S.~E.~Schneider, R.~Stiening, et al.,
    in {\it The Impact of Large Scale Near-IR Sky Surveys},
    Ed. by F.~Garzon et al.,
    (Kluwer Acad. Publ. Comp., Dordrecht 1997) p.25.
\bibitem{mass2b:Verkhodanov_n}
81.
 IPAC,
   Explanatory Supplement to the 2MASS Second Incremental Data Release,
   {\tt http://www.ipac.caltech.edu/2mass/releases/ \protect\linebreak 
/second/doc/explsup.html}
   (2002).
\bibitem{ifsc:Verkhodanov_n}
82.
M.~Moshir, IRAS Faint Source Survey, Explanatory supplement,
    Version 1 and tape,  Pasadena: Infrared Processing and Analysis
    Center, California Institute of Technology,  Ed. by M. Moshir 
(1989).
\bibitem{ipsc:Verkhodanov_n}
83.
 IRAS group, IRAS Point source catalogue; Catalogs and Atlases,
    Explanatory Supplement, Joint IRAS Science Working Group (1987).

\bibitem{c3opt:Verkhodanov_n}
84.
  H.~Spinrad, S.~Djorgovski, J.~Marr, and L.~Aguilar, \pasp~ {\bf 97}, 
       93 (1985).
\bibitem{txso:Verkhodanov_n}
85.
  E.~P.~Bozyan,
    \apjs~ {\bf 82},  1 (1992).
\bibitem{sdss:Verkhodanov_n}
86.
  D.~P.~Schneider, P.~B.~Hall, G.~T.~Richards, et al.,
    \aj~ {\bf 134}, 102 (2007).
\bibitem{vv11:Verkhodanov_n}
87.
  M.-P.~Veron-Cetty and P.~Veron, \aaa~ {\bf 412}, 399 (2003).
\bibitem{FaAPM:Verkhodanov_n}
88.
 R.~G.~McMahon, R.~L.~White, D.~J.~Helfand, et al.,
    \apjs~ {\bf 143}, 1 (2002).
\bibitem{ugc:Verkhodanov_n}
89.
W.~D.~Cotton, J.~J.~Condon, and E.~Arbizzani,
    \apjs~ {\bf 125}, 409 (1999).
\bibitem{pgc:Verkhodanov_n}
90.
G.~Paturel, P.~Fouque, L.~Bottinelli, and L.~Gouguenheim,
  \aas~ {\bf 80}, 299 (1989).
\bibitem{usno:Verkhodanov_n}
91.
D.~G.~Monet,
   The 526,280,881 Objects In The USNO-A2.0 Catalog,
    American. Asron. Soc., 19312003 (1998).
\bibitem{usno1:Verkhodanov_n}
92.
D.~Monet, B.~Canzian, H.~Harris, et al.,
   USNO-SA2.0,
  US Naval Observatory Flagstaff Station  (1997).


\bibitem{rosat:Verkhodanov_n}
93.
W.~Voges, B.~Aschenbach, Th.~Boller, et al.,
   \aaa~ {\bf 349}, 389 (1999).
\bibitem{wga:Verkhodanov_n}
94.
 N.~E.~White, P.~Giommi, and L.~Angelini,  \aas~ {\bf 185}, 4111 (1994).
\bibitem{einst:Verkhodanov_n}
95.
 E.~C.~Moran, D.~J.~Helfand, R.~H.~Becker and  R.~L.~White,
     \apj~ {\bf 461}, 127 (1996).

\bibitem{chandra:Verkhodanov_n}
96.
D.-W.~Kim,  R.~A.~Cameron, J.~J.~Drake, et al.,
  \apjs~ {\bf 150}, 19 (2004).
\end{thebibliography}
\end{document}